# SigViewer: Visualizing Multimodal Signals Stored in XDF (Extensible Data Format) Files*

Yida Lin[1], Clemens Brunner[2], Paul Sajda[1], *Fellow, IEEE,* and Josef Faller[1]

*Abstract*— Multimodal biosignal acquisition is facilitated by recently introduced software solutions such as Lab Streaming Layer (LSL) and its associated data format XDF (Extensible Data Format). However, there are no stand-alone applications that can visualize multimodal time series stored in XDF files. We extended SigViewer, an open source cross-platform Qt C++ application with the capability of loading, resampling, annotating, and visualizing signals stored in XDF files and successfully applied the tool for post-hoc visual verification of the accuracy of a system that aims to predict the phase of alpha oscillations within the electroencephalogram in real-time.

## I. INTRODUCTION

Multimodal biosignal acquisition, where different types of signals and events are recorded simultaneously, is important for both basic and applied biomedical research. A recently introduced software-based approach, Lab Streaming Layer (LSL) [1], allows ad-hoc distributed acquisition and synchronization of signals of diverse types, sampling rates, and meta information. The acquired and synchronized signals can be stored in an XDF [2] file using tools included in the LSL package (LabRecorder). To create a responsive stand-alone visualization application for multimodal biosignals, we extended the open source cross-platform Qt C++ application SigViewer [3] to now support the XDF format. Here we apply SigViewer for post-hoc verification of the accuracy of a system that determines the phase of a 10 Hz oscillation in a noisy signal to predict the phase 200 ms into the future so that an event (eventually a transcranial magnetic stimulation pulse) at a specific phase can be triggered.

## II. METHODS

To let SigViewer and other C++ applications load XDF files, we created a loading function for XDF and published it as a C++ library *libxdf* under the GPL [4]. For visualization, SigViewer can resample signals of different sampling rates to a common sampling rate using the open source library *Smarc*. The metadata of XDF files are stored in XML format and SigViewer can display those in a tree structure. SigViewer also automatically scales signals and computes the effective (actual) sampling rates based on the data and will notify the user if they deviate significantly from the nominal sampling rates. Event annotation can be important when viewing and analyzing biosignals, for example to indicate the presence of artifacts. We hence extended SigViewer with the functionality to add new events to XDF files. The added events can be added to the original file, or exported as a CSV file. We also redesigned the graphical user interface so that streams and channels can now be separately selected for visualization. Different streams and event types are visualized in contrasting colors.

The phase-prediction system here uses a simulated signal instead of actual electroencephalogram to test the algorithm. The system writes time series and events at different processing stages into LSL from where they are recorded into an XDF file. The XDF file is then visualized using SigViewer.

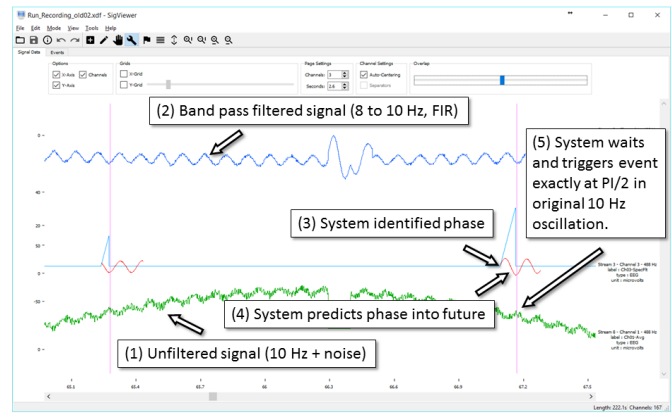

Figure 1: SigViewer allows detailed post-hoc inspection of multimodal signals and events written in different stages of our phase-prediction application. Text boxes and arrows were added for clarification. Numbers 1 to 5 represent the processing stages in the phase-prediction system.

## III. RESULTS AND DISCUSSION

In addition to the originally supported various biosignal file formats, SigViewer is now also able to visualize multi-stream XDF data in a fast and convenient user interface. In our tests, we used SigViewer successfully to verify post-hoc that a real-time phase-prediction system correctly triggered events at a specific phase. Together with the flexible distributed signal acquisition and synchronization framework LSL, SigViewer forms a complete open-source tool chain for multimodal signal acquisition and visualization.

*This work was funded by the Army Research Laboratory under cooperative agreement number W911NF-10-2-0022.

Y. Lin, P. Sajda and J. Faller are with LIINC at Columbia University, NYC, NY, USA. C. Brunner is with the Institute of Psychology, University of Graz, Austria. Correspondence: clemens.brunner@uni-graz.at.